\documentclass[prl,superscriptaddress,showpacs,twocolumn]{revtex4}

\usepackage{amsfonts}
\usepackage{amsmath}
\usepackage{graphicx}
\usepackage{amssymb}
\usepackage{amsmath}
\usepackage{amssymb}
\usepackage{graphicx}
\usepackage{subfigure}
\usepackage{lscape}
\usepackage{color}
\usepackage{epstopdf}

\begin{document}
\title{Simulating an interacting gauge theory with ultracold Bose gases}

\author{M.J.  Edmonds}
\affiliation{SUPA, Institute of Photonics and Quantum Sciences, Heriot-Watt University, Edinburgh EH14 4AS, United Kingdom}
\author{M. Valiente}
\affiliation{SUPA, Institute of Photonics and Quantum Sciences, Heriot-Watt University, Edinburgh EH14 4AS, United Kingdom}
\author{G. Juzeli\=unas}
\affiliation{Institute of Theoretical Physics and Astronomy, Vilnius University, A. Go\v{s}tauto 12, Vilnius 01108, Lithuania}
\author{L. Santos}
\affiliation{Institut f\"ur Theoretische Physik, Leibniz Universit\"at, Appelstrasse 2, D-30167 Hannover, Germany}
\author{P. \"Ohberg}
\affiliation{SUPA, Institute of Photonics and Quantum Sciences, Heriot-Watt University, Edinburgh EH14 4AS, United Kingdom}
\begin{abstract}
We show how density dependent gauge potentials can be induced in dilute gases of ultracold atoms using light-matter interactions. We study the effect of the resulting interacting gauge theory and show how it gives rise to novel topological states in the ultracold gas. We find in particular that the onset of persistent currents in a ring geometry is governed by a critical number of particles. The density-dependent gauge potential is also found to support chiral solitons in a quasi-one-dimensional ultracold Bose gas.  
 \end{abstract}
\pacs{03.75-b,03.75.Lm,03.65.Sq}
\maketitle

{\it Introduction}. 
Our understanding of the fundamental interactions between elementary particles is founded on gauge fields. The role of the gauge field is to mediate the interaction between particles. The simplest example we know of is electromagnetism where charged particles interact through exchanging virtual photons. The Coulomb potential between charged particles is encompassed by this gauge theory which can be recast in the familiar form of Maxwell's equations in the classical limit. Gauge theories are not restricted to electromagnetism only. The interactions in nuclei are governed by more complicated objects as far as gauge fields are concerned. There one has to use higher dimensions which typically requires a non-Abelian theory, such as the Yang-Mills field for the gluons \cite{qcd_book}. For all this to hold, the gauge fields must be dynamical. In other words  we must be allowed to construct a Lagrangian which also describes the propagation of the gauge field in vacuum. Solving the full quantum dynamics of such systems is a formidable task \cite{troyer_2005}. The solution could be to design a special purpose quantum simulator \cite{feynman_1982}.

{Very recently, the first few theoretical proposals in this direction have appeared \cite{banerjee_2012,Cirac,Zoller,Lewenstein}, where it was shown that it is in principle possible to simulate a dynamical gauge theory using cold atoms trapped in optical lattices. Also smaller steps towards the ambitious goal of simulating aspects of the standard model using possibly less demanding experimental techniques may provide some important insights (see for instance \cite{cirac_2010,keilmann_2011,dutta_2011})}. A more modest problem that generated intense interest in the late 1990s was the quest for finding a pure gauge theory with solutions given by the one-dimensional analog of the well-known two-dimensional anyons \cite{Wilczek}. The first attempt in this direction \cite{Benetton} failed to describe one-dimensional anyon solutions \cite{Benetton2}, but the associated semiclassical, non-linear model of the interacting gauge theory supported chiral solitons, as shown by Aglietti, Griguolo, Jackiw, Pi and Seminara (AGJPS) in Ref. \cite{aglietti_1996}. The generation of chiral solitons is clearly also an interesting goal to pursue in its own right due to the unconventional coherent transport mechanisms in the superfluid regime.

In this Letter, we show that under proper conditions conveniently engineered laser
fields similar to those employed in Refs. \cite{lin_2009a,lin_2009b,lin_2011} can induce an effective density-dependent vector potential in a weakly-interacting ultracold Bose gas, which constitutes the semiclassical limit of an interacting gauge theory for bosons. When the system is tightly confined such that it forms a quasi-one-dimensional gas, it is described, in a one-to-one fashion, by the AGJPS gauge theory \cite{aglietti_1996}. We show that the density-dependent gauge field leads to remarkable consequences, including density-dependent persistent currents in ring geometries, drifts in the free expansion dynamics, and chiral solitons in a Bose-Einstein condensate (BEC).

{\it An effective interacting gauge theory}. There are a number of ways to induce artificial magnetic fields in ultracold atomic gases ranging from stirring the cloud by a laser spoon or using asymmetric external traps \cite{madison_2000} to laser assisted tunneling in optical lattices which induce the required phases for the tunneling amplitudes between the different lattice sites \cite{jaksch_2003,aidelsburger_2011}. For ultracold atoms optically induced gauge potentials can also be created based on dark state dynamics \cite{dum_1996,visser_1998,juzeliunas_2004} or Raman transitions \cite{lin_2009a,lin_2009b,lin_2011}. These gauge potentials all have in common that they are static and given by the external rotation frequency or laser parameters; there is no dependence on the density of the atomic cloud in the gauge potential using these techniques. Here, we show how a density-dependent vector potential can arise in a weakly-interacting Bose-Einstein condensate based on collisionally induced meanfield shifts of the electronic levels in the atoms, which also constitutes the semiclassical limit of an interacting bosonic gauge theory. This can be done by considering a gas of optically addressed two-level atoms forming a BEC with internal state space given by $|1\rangle,|2\rangle$. 
Alkali atoms are usually good candidates for this, but fast spontaneous decay from these states might render them unusable, although a setup relying on dark states  \cite{dalibard_2011} could circumvent this problem. Alternatively, good candidates for experimentally realizing the effects discussed here would be for instance strontium, calcium or ytterbium, which can have radiative life times of several seconds \cite{ye_2008}.

The microscopic $N$-body Hamiltonian is given by 
\begin{equation}
\hat{H}=\sum_{q=1}^{N}\left(\frac{\mathbf{p}_q^2}{2m}+\hat{U}_q+\hat{V}_q\right)\otimes \hat{1}_{q}+\sum_{q<l=1}^{N}\hat{\mathcal{V}}_{q,l}\otimes \hat{1}_{q,l}, \label{microscopic}
\end{equation}
where
\begin{equation}
\hat{U}_q=\frac{\hbar\Omega}{2}\left( \begin{array}{cc}
0& e^{-i\phi_q}\\
e^{i\phi_q} & 0\\
\end{array}\right)
\label{lmc}
\end{equation}
is the Hamiltonian for the light-matter interaction and $\hat{V}_q$ is a single-particle external potential which we will in the following derivation put equal to zero for simplicity. It can readily be added to the resulting equation of motion if needed. The $\hat{1}_{q,\ldots}$ is the identity operator acting on the subspace excluding particles $q,\ldots$, whereas $\hat{\mathcal{V}}_{q,l} = \mathrm{diag}[g_{11},g_{12},g_{12},g_{22}]\delta(\mathbf{r}_q-\mathbf{r}_l)$ is a $4\times 4$ diagonal matrix describing the two-body interaction with coupling strengths $g_{ij} = 4\pi\hbar^2 a_{ij} /m$, with $a_{ij}$ the s-wave scattering length between the components $i$ and $j$.
In Eq. (\ref{lmc}), $\Omega$ is the two-photon Rabi frequency characterizing the light-matter coupling, $\phi_q\equiv \phi(\mathbf{r}_q)$ is the laser phase at particle $q$'s position, and the laser detuning from the atomic resonance is chosen to be zero for simplicity. However, the meanfield terms stemming from $\hat{\mathcal{V}}_{q,l}$ will introduce an effective detuning.
The corresponding Hamiltonian which takes into account collisional meanfield effects is then given by $\hat{H}_{GP}=\hat{p}^2/2m\otimes\hat{1}+\hat{\mathcal{V}}+\hat{U}$, where $\hat{\mathcal{V}}=(1/2){\rm diag}[g_{11}\rho_1+g_{12}\rho_2,g_{22}\rho_2+g_{12}\rho_1]$, with $\rho_i=|\Psi_i|^2$ ($i=1,2$) the density of population in the atomic state $i$, such that $\langle \hat{H} \rangle_{\Psi_{GP}} = \langle\Psi |\hat{H}_{GP}|\Psi\rangle$, where $|\Psi_{GP}\rangle=\otimes_{k=1}^{N}|\Psi_k\rangle$.

{For weakly interacting atoms}, the coupling strength $\hbar\Omega$ is typically much larger than the mean-field energies. The zero-th order approximation to the state of the system is chosen as the usual starting point in Bogoliubov's theory of the Bose gas. In this regime, to diagonalize $\hat{U}+\hat{\mathcal{V}}$ we treat $\hat {\mathcal{V}}$ as a small perturbation to $\hat U$. If we define the densities in the dressed states as $\rho_{\pm}=|\Psi_{\pm}|^2$, the corresponding eigenstates of $\hat U+\hat {\mathcal{V}}$ are given by the perturbed dressed states 
%\begin{equation}
$|\chi_{\pm}\rangle=|\chi_{\pm}^{(0)}\rangle+|\chi_{\pm}^{(1)}\rangle$,
%\end{equation}
where 
\begin{equation}
|\chi_{\pm}^{(1)}\rangle=\pm \frac{g_{11}-g_{22}}{8\hbar \Omega}\rho_{\pm}|\chi_{\mp}^{(0)}\rangle,
\end{equation}
with eigenvalues $g\rho_{\pm}\pm\hbar\Omega/2$, $g=(g_{11}+g_{22}+2g_{12})/4$ and $|\chi_{\pm}^{(0)}\rangle=(|1\rangle\pm e^{i\phi}|2\rangle)/\sqrt{2}$, together with the adiabatic approximation such that either $\rho_{-}({\bf r},t)\approx 0$ or $\rho_{+}({\bf r},t)\approx 0$. A general state can consequently be written like $|\varphi\rangle=\sum_{i=\{+,-\}}\Psi_{i}({\bf r},t)|\chi_i\rangle$. By projecting onto one of the dressed states, $|\chi_{\pm}\rangle$,  we obtain the effective Hamiltonian \cite{dalibard_2011}
\begin{equation}
\hat{H}_{\pm}=\frac{1}{2m}({\bf{p}}-{\bf{A}_{\pm}}[\mathbf{r};\rho_{\pm}(\mathbf{r},t)])^2+W\pm\frac{\hbar\Omega}{2}+\frac{g}{2}\rho_{\pm} 
\label{ham1}
\end{equation}
where $W=\frac{\hbar^2}{2m}|\langle \chi_-|\nabla \chi_+\rangle|^2$ is a scalar potential and ${\bf{A}}_{\pm}=i\hbar\langle\chi_{\pm}|\nabla\chi_{\pm}\rangle$ is a geometric vector potential that arises from the projection of the full system onto one of the dressed states.  In order for the adiabatic approximation to hold we must ensure that any induced detuning is small compared to the Rabi frequency $\Omega$.  The resulting vector potential is then given, to leading order, by 
\begin{equation}
{\bf{A}}_{\pm}={\bf A}^{(0)}\pm {\bf a}_1\rho_{\pm}({\bf r})\label{nonlin}
\end{equation}
where  ${\bf A}^{(0)}=-\frac{\hbar}{2}\nabla\phi$ is the single particle contribution and ${\bf a}_1=(\nabla\phi)(g_{11}-g_{22})/(8\Omega)$ controls the effective strength of the density-dependent vector potential.  

In order to derive a meanfield Gross-Pitaevskii type equation we apply the variational principle $\delta \mathcal{L}/\delta \Psi^* = 0$ to the action $\mathcal{L}=\langle\Psi |(i\hbar\partial_t-H_{\pm})|\Psi\rangle$, with respect to $\Psi^*$. We consider in the following the $+$ branch in $\hat H_\pm$ without loss of generality and consequently  drop the $\pm$ index in $\rho_{\pm}$, $\Psi_{\pm}$ and ${\bf A}_\pm$. The resulting equation of motion is then {
\begin{equation}
 \left[\frac{{(\bf{p}-\bf{A})}^2}{2m}+ {\mathbf a}_1\cdot {\mathbf j}+W+g\rho\right]\Psi=i\hbar \partial_t \Psi,\label{gp}
\end{equation}
where $\bf{A}$ is given by Eq. \eqref{nonlin}} together with a nonlinearity in the form of a current,
\begin{equation}
{\bf j}=\frac{\hbar}{2mi}\left[\Psi\left(\nabla +\frac{i}{\hbar}{\bf A}\right)\Psi^*-\Psi^*\left(\nabla -\frac{i}{\hbar}{\bf A}\right)\Psi\right].
\end{equation}
The meanfield scalar potential $W$ is given to leading order by $W={|{\mathbf{A}^{(0)}}|^2}/{2m}$.

{\it One-dimensional physics}. The density-dependent vector potential gives rise to a number of interesting and counterintuitive scenarios. To illustrate this we will in the following assume that the cloud of atoms is tightly confined such that any motion in the transversal direction is frozen out and the dynamics is well described by an effectively one dimensional meanfield description. We choose $\phi=kx$ as the phase of the incident laser, together with the transformation $\Psi(x)=e^{-ikx/2} \psi(x)$, which results in the equation 
\begin{equation}
i\hbar\partial_t \psi=\left[\frac{1}{2m}\left({\hat p}- a_1\rho\right)^2+ a_1j+\tilde{W}+g\rho\right]\psi,\label{schro}
\end{equation}
where $\tilde{W}=\hbar^2k^2/8m$, and $a_1=k(g_{11}-g_{22})/8\Omega S_t$ characterises the strength of the current nonlinearity. The effective transversal area of the 1D cloud is given by $S_t$. Our model is found to be equivalent to the AGJPS model \cite{aglietti_1996}, with the additional non-linear interaction term $g\rho$.

%%%%%%%%%

The current $a_1j(x)$ can be made influential provided that the meanfield shift is relatively large. 
The combination of the three parameters $\Omega,\rho$ and $g_{11}-g_{22}$ in $a_1$ allows for great flexibility in tuning the strength of the gauge field. For instance, with a density of $6.0 \times10^{14}$ cm$^{-3}$, a difference in scattering lengths $a_{11}-a_{22}=5.0$ nm using for instance optical Feshbach resonances \cite{fedichev_1996,theis_2004,enomoto_2008,papoular_2010}, and a Rabi frequency of $185$ kHz, one obtains the ratio $(g_{11}-g_{22})\rho/\hbar\Omega=0.01$ which can affect the dynamics (see also Figs. 1 and 2). It should be noted that for standard BEC setups such as $^{87}$Rb, this parameter would be vanishingly small due to the small difference between the scattering lengths. However, by carefully tuning the parameters one can circumvent such problems, as illustrated above.
In the following we will study three scenarios which illustrate the role of the density dependent gauge field.

%%%%%%%%%%%

{\it Density dependent persistent currents.} We consider at this point a 1D ring-like geometry in the x-y plane and an additional laser beam propagating in the z-direction which carries an orbital angular momentum with $\phi=\ell\theta$ where $\ell$ is an integer.  This configuration gives rise to a gauge potential in the azimuthal $\theta$-direction, hence the situation is similar to the linear 1D case, but now with periodic boundary conditions. The time-independent Gross-Pitaevskii equation on the ring of radius $R$ is obtained from equation (\ref{schro}) by setting $x=R\theta$ and $\psi(x,t)=\psi(x)\exp[-iEt/\hbar]$.
The solutions are given by $\psi(\theta)=\sqrt{\frac{N}{2\pi R}}e^{iq\theta}$ with normalisation condition $R\int_{0}^{2\pi}d\theta|\xi(\theta)|^2=N$, where $N$ is the number of particles in the ring. The energy difference between two different angular momentum states can readily be calculated,
\begin{equation}
E_{q+1}-E_q=\frac{1}{2m}\left[\frac{2\hbar}{R}\left(\frac{\hbar q}{R}- a_1\rho\right)+\frac{\hbar^2}{R^2}\right],
\label{rot}
\end{equation}
where $q$ is an integer number which labels the quantized rotation of the cloud. We see from Eq. \eqref{rot} that the ground state configuration becomes a function of the number of particles. Interestingly, this implies that at a certain critical density,
\begin{equation}
\rho_c(q)=\frac{8\hbar \Omega}{\ell (g_{11}-g_{22})}(q+1/2),
\end{equation}
the ground state changes from one rotational state to another with $q\rightarrow q+1$. This is in contrast to the standard situation for a ring BEC under rotation, where the onset of a current is given by the rotation frequency.

\begin{figure}
\includegraphics[scale=0.5]{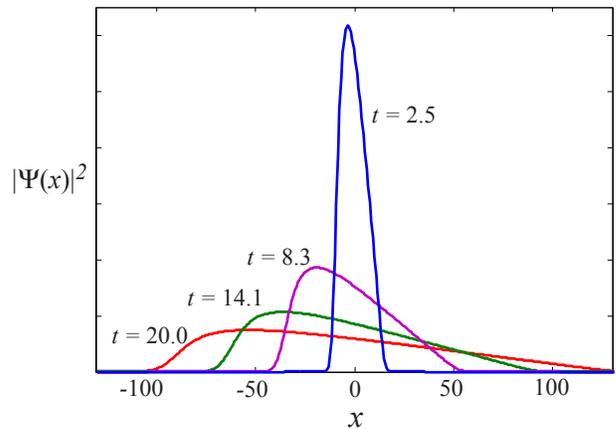}
\caption{\label{fig1} (Color online) Snapshots of the free expansion of a harmonically trapped BEC with $(gN/S_t) (2m/\hbar^2k)=30$ and trap frequency $\omega_t=\hbar k^2/2m$. The expansion is asymmetric where a change of sign in $a_1$ changes the direction of the drift. The coupling strength for the gauge field was $(g_{11}-g_{22})kN/(S_t \hbar\Omega)=5$. Length is in units of $1/k$ and time in units of $2m/\hbar k^2$.}
\end{figure}

{\it Free expansion drift}. A numerical solution of Eq. \eqref{schro} shows that the free expansion is no longer symmetric (see Fig. \ref{fig1}). In addition the current term induces a drift which is proportional to $a_1$ times the density of the BEC. The onset of a drift can be explained using as variational ansatz the solution of a freely expanding wavepacket where we allow for a drift velocity $\dot x_0$ of the centre of mass, 
\begin{equation}
\phi(x,t)=\bigg(\frac{N^2}{\pi\sigma_x(t)^2}\bigg)^{1/4}\exp\bigg(-\frac{(x-x_0(t))^2}{2\sigma_x(t)^2}\bigg)e^{i\mathcal{S}}.
\label{var1}
\end{equation}   
The spatially varying phase is given by $\mathcal{S}=m\dot{x}_0(x-x_0)/\hbar$,  $\sigma_x(t)=\sigma_0\sqrt{1+t^2/\tau^2}$ with $\tau=2m/k^2\hbar$ is the time dependent width of the gaussian and $N$ is the number of particles. From Eq. \eqref{var1} and the corresponding Lagrangian density
we obtain an equation of motion for the position $x_0(t)$ of the wave packet,
\begin{equation}
m\ddot{x}_{0}=\frac{\sqrt{2}a_1 N\dot{\sigma}_x(t)}{\sqrt{\pi}\sigma_x(t)^2}.
\label{eq_mo}
\end{equation}
Equation \eqref{eq_mo} and its solution 
\begin{equation}
x_0(t)=\bigg(\frac{\sqrt{2}a_1 N\tau}{\sigma_0\sqrt{\pi}m}\bigg)\bigg[\frac{t}{\tau}-{\rm arcsinh}(t/\tau)\bigg],
\label{sol}
\end{equation}
gives us a way to understand the effect of the dynamical gauge potential on the condensate as it expands. The increasing width as a function of time drives the drift of the center of mass coordinate $x_0$.

Figure \ref{fig1} shows that the presence of the current term in \eqref{schro} causes the free expansion of the condensate to experience a drift that depends on both the sign and magnitude of the strength of the dynamical gauge field captured by the parameter $a_1$. The onset of a drift can also be understood as an effect of the asymmetric coupling of the different momentum components in the initial wave packet with the density of the cloud.

%%%%%%%%%%%%%%%%%%%%%%%%%%%

{\it Chiral solitons.} Our semiclassical gauge theory, whose equation of motion is given by Eq. (\ref{schro}), supports chiral solitons. The existence of chiral soliton solutions is also ultimately a consequence of the breakdown of Galilean relativity in the corresponding microscopic version of our gauge theory.
  
We begin by using the gauge transformation,
\begin{equation}
\psi(x,t)=\exp\bigg(\frac{ia_1}{\hbar}\int\limits^{x}_{-\infty} dx'\rho(x',t)-i\tilde W t/\hbar\bigg)\Phi(x,t).
\end{equation}
Eq. \eqref{schro} then simplifies to
\begin{equation}
i\hbar\partial_t{\Phi}=\bigg[-\frac{\hbar^2}{2m}{\partial_x^2}-2a_1 j(x)+g|\Phi|^2\bigg]\Phi,
\label{schro2}
\end{equation}
where the gauge-transformed current becomes
\begin{equation}
j(x)=\frac{\hbar}{2mi}\bigg[\Phi^{*}(x)\partial_x\Phi(x)-\Phi(x)\partial_x\Phi^{*}(x)\bigg].
\end{equation}
Equation \eqref{schro2} can be solved by first writing the wave function in the form \cite{harikumar_1998}
\begin{equation}
\Phi(x,t)=\xi(x-ut)e^{i(umx-(\frac{1}{2}mu^2+\mu t))/\hbar},
\end{equation}
where $\xi(x-ut)$ is a real valued function and $\mu$ is the chemical potential.  The current consequently transforms into $j(x)=u\xi^2$ where $u$ is the speed of the soliton. The resulting differential equation for $\xi(x-ut)$ is
\begin{equation}
\mu\xi=-\frac{\hbar^2}{2m}\partial_x^2\xi+(g-2a_1u)\xi^3.
\label{nlse}
\end{equation}
For $\tilde g=g-2a_1u>0$ we find in particular the dark soliton solution 
\begin{equation}
\frac{\Phi(x,t)}{\sqrt{\rho_0}}=\frac{\exp[i(mu(x-ut)-(\frac{1}{2} mu^2+\mu) t)/\hbar]}{\coth[(x-ut)/(\sqrt{2}l_0)]}
\label{ds}
\end{equation}
where $\rho_0$ is the background density, $\mu=\tilde g\rho_0$ and $l_0=\hbar/\sqrt{2m\tilde g\rho_0}$.
For $\tilde g<0$ we obtain a bright soliton 
\begin{equation}
\frac{\Phi(x,t)}{\sqrt{\rho_0}}=\frac{\exp[i(mu(x-ut)-(\frac{1}{2} mu^2-\mu) t)/\hbar]}{\cosh((x-ut)/(\sqrt{2}l_0)]}
\label{bs}
\end{equation}
with $\mu=|\tilde g|\rho_0/2$.
The solutions in equation \eqref{ds} and \eqref{bs} are chiral, which means that the solitons can only propagate in a specific direction for a chosen velocity.  Interestingly, if $g=2a_1u$ we are in a situation where the current non-linearity cancels the mean field interactions between particles, with no soliton solutions present.
Depending on the precise physical setup this particular situation may or may not be possible to reach due to a breakdown of the adiabatic assumption or a violation of the perturbative assumption. 

\begin{figure}
\includegraphics[scale=0.48]{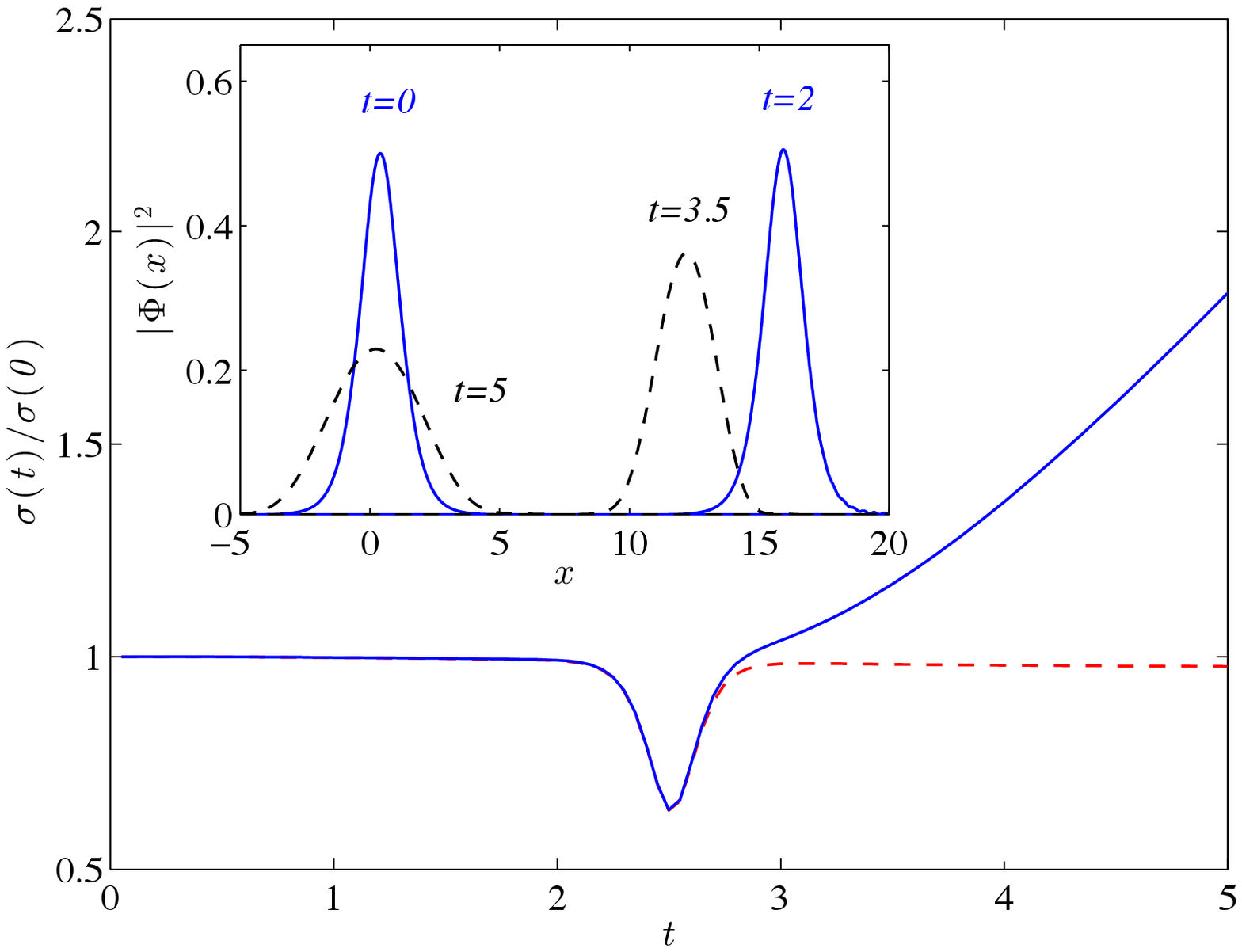}
\caption{\label{ref} (Color online) The normalised width $\sigma(t)/\sigma(0)$ of the bright soliton. The blue solid line shows $\sigma(t)/\sigma(0)$ with current strength $(g_{11}-g_{22})kN/(S_t \hbar\Omega)=0.125$ and $(gN/S_t) (2m/\hbar^2k)=-0.5$, whilst the red dashed line indicates the width of an initially identical soliton without the current nonlinearity ($a_1=0$). After reflection the soliton starts expanding due to the change in nonlinear strength. The inset shows snapshots of the density of the soliton prior to (solid blue) and after reflection (dashed black) at times $t=0,2.0,3.5,5.0$. All lengths are in units of $1/k$ and time in units of $2m/\hbar k^2$.}
\end{figure}

The concept of a chiral soliton can be illustrated by considering the reflection of a BEC from a hard wall. In Fig. \ref{ref} we show how a bright soliton initially moving in the positive x-direction is destroyed after reflection. A standard bright soliton would retain its width $\sigma(t)=\sqrt{\langle x^2\rangle}$ after reflection whereas the chiral soliton is found to start to expand after reflection. The change in the nonlinear strength due to the change in momentum after the reflection results in a state which is not the soliton solution any more, hence the solution is no longer confined.

{\it Conclusions.} In this Letter we have shown how an interacting gauge theory for a BEC can be generated. 
The resulting gauge field is not fully dynamical, in the sense that it is always zero if no matter field is present. The emerging gauge field does however depend on the density of the BEC, and therefore constitutes an interacting field with a back-action between the BEC dynamics and the gauge field.
The equation of motion includes a current non-linearity and in the quasi-one-dimensional regime our model is identical to the Aglietti-Griguolo-Jackiw-Pi-Seminara gauge theory \cite{aglietti_1996}. The coupling of the BEC to its current gives rise to a number of exotic types of dynamics. We have shown how the presence of topological states corresponding to persistent currents in a ring geometry depend on the number of particles. Also soliton solutions can be identified which are chiral in nature. It is certainly tempting to draw analogies between the atomic system considered here and models of field theories describing the fundamental forces between elementary particles.  From a quantum simulator point of view, perhaps the most intriguing aspect would be a generalisation of the mechanisms discussed here to a pseudo-spin situation which can also support non-Abelian gauge potentials \cite{osterloh_2005,ruseckas_2005}.

\acknowledgements{We acknowledge helpful discussions with Jonas Larson, Maciej Lewenstein and Ian Spielman. M.J.E. acknowledges support from the EPSRC CM-DTC, M.V. and P.\"O.  from EPSRC grant No. EP/J001392/1, G.J. from grant No. MIP-082 by the Lithuanian Research Council, and L.S. from  the Center of Excellence QUEST.}

\bibliographystyle{unsrt}

\end{document}